\documentclass[amsmath,amssymb,aps,prd,preprint,superscriptaddress,nofootinbib]{revtex4-1}

\usepackage{braket}
\usepackage{graphicx}
\usepackage{grffile}
\usepackage{listings}

\newcommand{\beq}{\begin{eqnarray}}
\newcommand{\eeq}{\end{eqnarray}}
\newcommand{\rp}{\right)}
\newcommand{\lp}{\left(}

\newcommand{\rbb}{\right]} 
\newcommand{\lbb}{\left[}
\newcommand{\del}{\partial}
\newcommand{\tp}{t^\prime}
\newcommand{\tpp}{t^{\prime\prime}}
\newcommand{\vx}{\vec{x}}
\newcommand{\vy}{\vec{y}}
\newcommand{\vp}{\vec{p}}
\newcommand{\vq}{\vec{q}}
\newcommand{\vk}{\vec{k}}

\newcommand{\half}{\frac{1}{2}}
\DeclareMathOperator{\Tr}{Tr}

\DeclareMathOperator{\sgn}{sgn}

\usepackage{color}
\usepackage[normalem]{ulem}

\renewcommand\sout{\bgroup \color{red} \ULdepth=-.5ex \ULset}

\newif\iffigure
\figuretrue

\begin{document}

\preprint{YITP-17-49}

\title{Thermalization of overpopulated systems 
in the 2PI formalism  }

\author{Shoichiro Tsutsui}
\affiliation{
  Department of physics, Kyoto University, Kyoto 606-8502, Japan}

\author{Jean-Paul Blaizot}
\affiliation{
Institut de Physique Th\'eorique, CNRS/UMR 3681, CEA Saclay, F-91191 Gif-sur-Yvette, France }
\affiliation{
  Yukawa Institute for Theoretical Physics, Kyoto University, Kyoto 606-8502, Japan}%

\author{Yoshitaka Hatta}
\affiliation{
  Yukawa Institute for Theoretical Physics, Kyoto University, Kyoto 606-8502, Japan}%

\date{\today}

\begin{abstract}
\vspace{4mm}
Based on the two-particle irreducible (2PI) formalism to next-to-leading order in the $1/N$ expansion, we study the thermalization of overpopulated systems in scalar $O(N)$ theories with moderate coupling. We focus in particular on the growth of soft modes, and examine whether this can lead to the formation of a transient Bose-Einstein condensate (BEC)  when the initial occupancy is high enough. For the value of the coupling constant used in our simulations, we find that while the system rapidly approaches the condensation threshold, the formation of a BEC is eventually hindered by particle number changing processes.   
\end{abstract}

\pacs{Valid PACS appear here}
\maketitle

\section{Introduction}
Understanding the thermalization process of relativistic heavy-ion collisions remains an outstanding issue in the physics of quark-gluon plasmas (see a recent review \cite{Gelis:2016upa} and references therein). It is commonly assumed that the initial state of the matter formed after a collision is dominated by gluons whose typical momentum is the saturation scale $Q_s$,  and whose occupation number $f$ is inversely proportional to the strong coupling constant $\alpha_s$.
In high energy collisions, such as those at the Large Hadron Collider (LHC), the occupation number may become large,  $f\sim 1/\alpha_s(Q_s) \gg 1$, leading to an \emph{overpopulated} initial state. By this we mean that, given their total energy, the number of the initial gluons, if conserved during the evolution, is too large to be accommodated in the final state in a thermal distribution . 

In general, when overpopulated systems are driven towards  thermal equilibrium by elastic processes which conserve particle number, the particles in the final state that cannot be accommodated in the Bose-Einstein distribution populate a  Bose-Einstein condensate (BEC). The formation of such a condensate has been much studied in various contexts, mainly within kinetic theory (see \cite{Semikoz:1994zp,Semikoz:1995rd,Lacaze:2001qf,Blaizot:2013lga,Epelbaum:2014mfa,Xu:2014ega,Meistrenko:2015mda}, and references therein for some representative works).
Most of these works only consider elastic $2\to 2$ scattering in the Boltzmann equation.    Inelastic scattering may qualitatively change the picture \cite{Blaizot:2011xf,Kurkela:2011ti}, allowing only for the formation of  a transient BEC  \cite{Blaizot:2011xf}. A recent analysis of the specific effect of $2\to 3$ processes, dominated in QCD by collinear splittings, suggests a complete hindrance of BEC formation \cite{Blaizot:2016iir} . 

The validity of kinetic theory to discuss this kind of phenomenon may be questioned. It  is known in particular \cite{Berges:2004yj}, from the way it can be derived from the field equations of motion, that the Boltzmann equation is only an approximate description of the full nonequilibrium evolution, that is best suited for the dilute regime $f(p) \ll 1/\lambda$ where $\lambda\ll 1$ is the coupling constant. Strictly speaking then, the soft momentum regime $f(p\approx 0) \gg 1/\lambda$, which is important for the discussion of the BEC formation, is outside the region of applicability of the Boltzmann equation. While the range of applicability of kinetic equations may be wider than what is suggested by a specific derivation from microscopic physics, it is interesting to explore the dynamics of BEC formation using more complete formalisms. This has been done for instance within   classical statistical field theory in \cite{Berges:2012us} (see also \cite{Berges:2012ev,Berges:2014bba}), where the  formation of a transient BEC has been observed, as well as its subsequent  decay due to the inelastic processes which are naturally included in this formalism. However, the classical statistical field theory is sensitive to the ultraviolet cutoff, unless the coupling constant is chosen extremely small, $\lambda \ll 1$,  corresponding to very large occupations $f\gg 1$ \cite{Epelbaum:2014mfa}. If one has in mind getting insight into the situation in QCD, where the coupling $\alpha_s \sim 0.3 - 0.4$ is not particularly weak, we should explore other regimes.


In this paper, we investigate the problem of the BEC formation in the two-particle irreducible (2PI) formalism in the scalar $O(N)$ theory to next-to-leading order in the $1/N$ expansion \cite{Berges:2001fi,Aarts:2001yn,Aarts:2002dj}. This  framework allows us to explore the entire momentum regime, from the deep infrared where $f(p)\gg 1/\lambda$ all the way to the ultraviolet regime where $f$ exhibits the exponential Boltzmann tail. Moreover, the calculations are not limited to weak coupling, and allow for the study of arbitrary occupations. Unfortunately, in practice, these calculations involve heavy numerics.  As a result simulations have  been  sparse in the literature, often done in 1+1 or 1+2 dimensions \cite{Berges:2001fi,Aarts:2001yn,Nishiyama:2010wc,Hatta:2012gq}, and in 1+3 dimensions on small lattices \cite{Arrizabalaga:2004iw,Tranberg:2008ae}. 
 Very recently, a large volume simulation  in 1+3 dimensions has appeared \cite{Berges:2016nru}, but the focus there was on the so-called non-thermal fixed points and the self-similar behaviours which emerge far-from-equilibrium at   small coupling. Instead, we shall work at moderate coupling, so that the system eventually approaches equilibrium in a finite (computer) time, and we focus on whether and how a BEC is formed as we dial the initial occupancy. We shall indeed find that the system approaches condensation for large enough initial occupancies. However, our choice of a relatively large coupling enhances the effect of  inelastic processes which eventually hinder the condensation. 
  


\section{Time evolution equation}\label{Sec:evolution equation}

The 2PI formalism for non equilibrium dynamics that we shall use in this paper is well developed and extensively discussed in the literature (for a pedagogical introduction see e.g. \cite{Berges:2004yj}). In this section, we just summarize the main equations that we solve. 
We consider a scalar field theory with $O(N)$ symmetry in a non-expanding geometry.
The action  is given by
\begin{align}
	S[\varphi] = \int d^4x 
	\lbb 
	\half \del_\mu \varphi_a \del^\mu \varphi_a 
	- \half m^2 \varphi_a \varphi_a
	- \frac{\lambda}{4! N} (\varphi_a \varphi_a)^2
	\rbb,
	\label{classical action}
\end{align}
where $a = 1, \dots, N$.
In the 2PI formalism, the basic building blocks of the non-equilibrium evolution equations are the statistical function 
\begin{align}
F_{ab}(x,y)=\langle \{ \varphi_a(x),\varphi_b(y)\}\rangle,
\end{align}
and the spectral function 
\begin{align}
\rho_{ab}(x,y) = i\langle [\varphi_a(x),\varphi_b(y)]\rangle,
\end{align}
where $\{,\}$ and $[,]$ denote respectively an anticommutator and a commutator. 
We assume unbroken $O(N)$ symmetry so that the one-point function $\langle \varphi_a(x)\rangle$ vanishes at all times,  and the 2-point functions take the forms $F_{ab}=\delta_{ab}\,F$, $\rho_{ab}=\delta_{ab}\,\rho$. We also assume that the system is spatially homogeneous and isotropic, and perform a Fourier transformation to momentum space
%
%
\begin{align}
F(x,y) = \int \frac{d^3 p}{(2 \pi)^3} F(t,t',p) e^{i \vp \cdot (\vx - \vy) } .\qquad \rho(x,y) = \int \frac{d^3 p}{(2 \pi)^3} \rho(t,t',p) e^{i \vp \cdot (\vx - \vy) } ,
\end{align}
where $F(t,t',p)$ depends only on the modulus $p=|\vec{p}|$ of the momentum, and similarly for $\rho(t,t',p)$.
The evolution equations  then read 
\begin{align}
\lp \del_t^2 + p^2 + M^2(t) \rp F(t,\tp,p) 
=
&-\int^t_{t_0} d\tpp \Sigma_\rho(t,\tpp,p) F(\tpp,\tp,p) \notag \\
& \qquad+\int^{\tp}_{t_0} d\tpp \Sigma_F(t,\tpp,p) \rho(\tpp,\tp,p) , \label{evolF} \\
\lp \del_t^2 + p^2 + M^2(t) \rp \rho(t,\tp,p) 
= 
&-\int^t_{\tp} d\tpp \Sigma_\rho(t,\tpp,p) \rho(\tpp,\tp,p), \label{evolR}
\end{align}
%
%
%
where $M$ is the effective  mass (the local part of the self energy)
\begin{align}
M^2(t)
&=
m^2 + \frac{\lambda(N+2)}{6N} \int \frac{d^3 p}{(2\pi)^3} F(t,t,p). \label{p}
\end{align}
The right hand sides of  Eqs.~\eqref{evolF} and~\eqref{evolR} contain memory effect from the initial time $t_0=0$ to the current time $t$ and $\tp$.

\begin{figure}[hbt]
 \includegraphics[width=15cm]{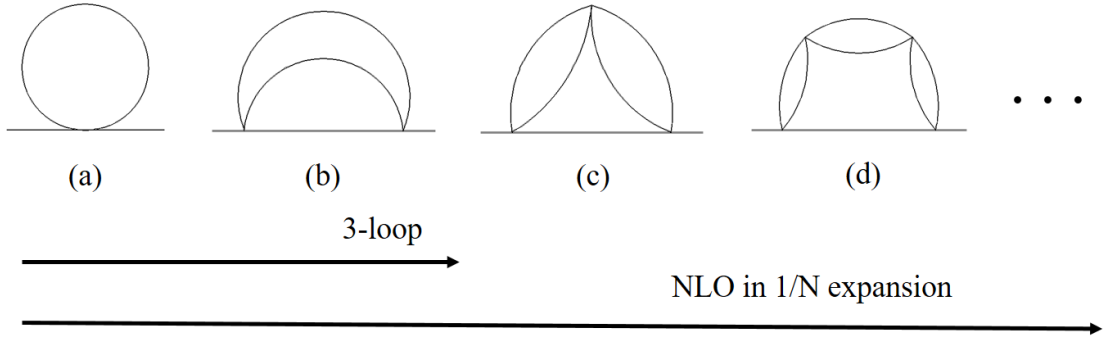}
\caption{The ring diagrams resummed in the loop and $1/N$ expansions. The first diagram (tadpole) corresponds to the mass shift (\ref{p}). The `3-loop approximation' which we shall discuss later on includes only the first two diagrams (a) and (b) in this figure. The sum of the NLO diagrams can be partially understood as a correction to the 3-loop diagram where one of the two vertices is replaced by a \textit{screened} interaction that corresponds to the sum of the bubble diagrams.
	} \label{diagrams} 
\end{figure}

The self energies $\Sigma_{F}$ and $\Sigma_\rho$ are given by the sum of 2PI diagrams starting from three loops. We shall however go beyond the skeleton loop expansion, and use here the 
 large-$N$ approximation,  and resum the `ring diagrams' which appear at the next-to-leading order in the $1/N$ expansion \cite{Aarts:2002dj}.
 In this approximation, the self energies are evaluated as 
\begin{align}\label{Sigma}
\Sigma_F(t,\tp,p)
&=
- \frac{\lambda}{3N} \int \frac{d^3 k}{(2\pi)^3} 
\lp 
F(t,\tp,|\vp-\vk|) I_F(t,\tp,k) - \frac{1}{4} \rho(t,\tp,|\vp-\vk|) I_\rho(t,\tp,k)
\rp ,\\
\Sigma_\rho(t,\tp,p)
&=
- \frac{\lambda}{3N} \int \frac{d^3 k}{(2\pi)^3} 
\lp 
F(t,\tp,|\vp-\vk|) I_\rho(t,\tp,k) + \rho(t,\tp,|\vp-\vk|) I_F(t,\tp,k)
\rp ,
\end{align}
where
\begin{align}
I_F(t,\tp,k)
&=
\frac{\lambda}{6} \int \frac{d^3 q}{(2\pi)^3} 
\Big[
F(t,\tp,|\vk-\vq|)F(t,\tp,q)   - \frac{1}{4} \rho(t,\tp,|\vk-\vq|)\rho(t,\tp,q)  \label{if} \\
& \qquad -\int^t_{t_0} d\tpp I_\rho(t,\tpp,k) 
\lp
F(\tpp,\tp,|\vk-\vq|)F(\tpp,\tp,q)  - \frac{1}{4} \rho(\tpp,\tp,|\vk-\vq|)\rho(\tpp,\tp,q) 
\rp \notag\\
&\qquad +2\int^{\tp}_{t_0} d\tpp I_F(t,\tpp,k) \rho(\tpp,\tp,|\vk-\vq|)F(\tpp,\tp,q) 
\Big] ,\notag\\
I_\rho(t,\tp,k)
&=
\frac{\lambda}{3} \int \frac{d^3 q}{(2\pi)^3}
\Big[
\rho(t,\tp,|\vk-\vq|)F(t,\tp,q) \label{irho} \\
& \qquad -\int^{t}_{\tp} d\tpp I_\rho(t,\tpp,k) \rho(\tpp,\tp,|\vk-\vq|)F(\tpp,\tp,q)  \notag
\Big] . 
\end{align}
%
%
The diagrams resummed in this expansion scheme are shown in Fig.~\ref{diagrams}. 

In thermal equilibrium, the statistical and spectral functions take the form:
\begin{align}
F_\text{eq}(t, \tp, p) &= \frac{1}{\omega_p} \lp \half + f_\text{BE}(\omega_p) \rp \cos \omega_p(t-\tp), 
\label{F in equilibrium}\\
\rho_\text{eq}(t, \tp, p) &= \frac{1}{\omega_p}  \sin \omega_{p}(t-\tp) , \label{bose}
\end{align}
where $\omega_p$ is the particle energy and $f_\text{BE}(\omega_p) = 1/(e^{\beta\omega_p} - 1 )$ is the Bose-Einstein distribution. Since the particle number is not conserved, the chemical potential $\mu$ vanishes in equilibrium.
The memory integrals in Eqs.~\eqref{evolF} and~\eqref{evolR} vanish for these distributions in the long time limit $t, \tp \to \infty$. 

When the system is not in equilibrium, but the quasi-particle picture is valid (which occurs on short time scales), one may define the quasi-particle distribution and the quasi-particle energy from the following equations \cite{Berges:2004yj}
\begin{align}
f(t,p) &= \sqrt{ F(t, \tp, p) \del_t \del_{\tp} F(t, \tp, p) - \del_t F(t, \tp, p) \del_{\tp} F(t, \tp, p)} \Big|_{t=\tp} - \half \,, \\
\omega_p(t) &= \sqrt{ \frac{\del_t \del_{\tp} F(t, \tp, p)}{F(t, \tp, p)}}\Bigg|_{t=\tp}, \quad  m_\text{qp}(t) = \omega_{p=0}(t), \label{mass}
\end{align}
where $m_\text{qp}$ is the quasi-particle mass.

Motivated by the form of the equilibrium 2-point function (\ref{F in equilibrium}), we choose the following initial conditions for Eqs.~(\ref{evolF}) and (\ref{evolR}) 
\begin{align}
	F(t, \tp, p) \big|_{t=\tp=0} &= \frac{1}{\sqrt{p^2 + m^2}} \lp \half + f_\text{in}(p) \rp, \\
	\del_t F(t, \tp, p) \big|_{t=\tp=0} &= 0, \\
	\del_t \del_{\tp}F(t, \tp, p) \big|_{t=\tp=0} &= \sqrt{p^2 + m^2} \lp \half + f_\text{in}(p) \rp,  \label{initial PD}
\end{align}
where 
\begin{align}
	f_\text{in}(p) &= A \theta( Q-p).
\end{align}
The initial conditions for the spectral function are automatically fixed by the equal-time commutation relation as  $\rho(t,\tp)|_{t=\tp=0}=0$, $\del_t\rho(t,\tp)|_{t=\tp=0}=1$ and $\del_t \del_{\tp}\rho(t,\tp)|_{t=\tp=0}=0$. 
 The momentum $Q$ is the analog  of the `saturation momentum' in the context of heavy-ion collisions.  
The parameter $A$ characterizes the initial occupancy. We shall be interested in how the nature of the thermalization process changes as we dial $A$ to larger and larger values.   A general expectation is that for large enough values of $A$, $A>A_c$, a Bose-Einstein condensate is formed. A crude estimate of the critical value $A_\text{crit}$ is  as follows \cite{Blaizot:2011xf}. Assuming $m=0$ for simplicity, the initial number density and the energy density are
\beq
n_\text{in}= \frac{AQ^3}{6\pi^2}\,, \qquad \varepsilon_\text{in}= \frac{AQ^4}{8\pi^2}\,,
\eeq
per degree of freedom. 
If the system eventually thermalizes at the temperature $T$ and vanishing chemical potential, we have, for a noninteracting theory,
\beq
n_\text{eq}=\frac{\zeta(3)}{\pi^2}T^3\,, \qquad \varepsilon_\text{eq}=\frac{\pi^2}{30}T^4\,.
\eeq
Condensation occurs if the dimensionless ratio $n_\text{in}/\varepsilon_\text{in}^{3/4}$ exceeds the equilibrium value $n_\text{eq}/\varepsilon_\text{eq}^{3/4}$. This leads to the condition $A\ge A_\text{c}\approx 0.15$. It is straightforward to generalize this argument to the case of non interacting massive particles (see also \cite{Meistrenko:2015mda}). In Fig.~\ref{ac}, we plot the function $A_\text{c}(m)$ which we obtained numerically.  Note that this estimate is only an approximation in the case of the fully interacting theory:  it ignores indeed the effects of the interactions, aside from that of giving a mass to the excitations.

\iffigure
\begin{figure}[hbt]
 \includegraphics[width=8cm]{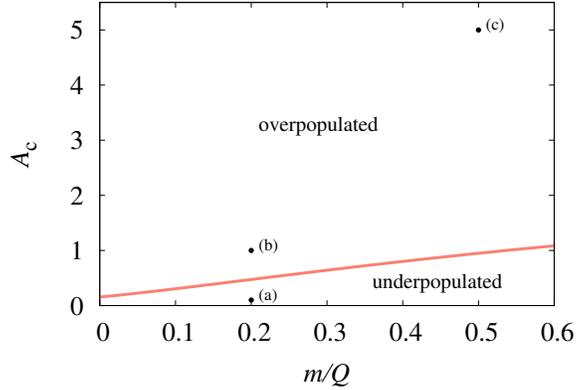}
\caption{ The critical initial occupancy $A_\text{c}$ as a function of the mass.
	Three points (a)-(c) denote the parameter sets that we employ in our numerical simulations. 
	} \label{ac} 
\end{figure}
\fi

It is interesting to study whether and how the  expectation of BEC formation is borne out in our field theoretic system where the coupling is not too small and the particle number is not conserved.  
To quantify the latter effect, we follow the evolution in time of the number density $n(t)$,  and the time derivative of its integral within a sphere of radius $p$, $J(t,p)$:
\begin{eqnarray}
n(t) &=& \int \frac{d^3 p}{(2\pi)^3} f(t,p) = \frac{1}{2\pi^2} \int_0^\infty dp p^2 f(t,p)\,, \label{number} \\
&& J(t,p) \equiv \frac{d}{dt} \int^p_0 d^3p' f(t,p')\,. \label{flux}
\end{eqnarray}
The potential delta-function contribution at $p=0$ is kept in $n(t)$ when we discretize the momentum integral to a sum over different $p$-bins.\footnote{We do not introduce a specific field to describe the potential condensate, as done e.g in Ref.~\cite{Berges:2012us}, Eq.~(4). } If the particle number is conserved, and if no particle accumulates in the state $p=0$, $J$ can be interpreted (to within a factor $4\pi p^2$) as the flux of particles in momentum space through the sphere of radius $p$ (counted positively if the particles are  towards $p=0$). In the present situation, because of the existence of inelastic processes that do not conserve particle number, $J$ also includes, in addition to the flux just mentioned, contributions from particle creation and annihilation inside the sphere of radius $p$. 

We also control the time evolution of the total energy density, given by
\begin{align}
\varepsilon(t) 
&= 
\frac{1}{2}\int \frac{d^3p}{(2\pi)^3}[\partial_t \partial_{t'}+p^2+m^2] F(t,t',p)|_{t=t'} 
+ 
\frac{\lambda(N+2)}{24N}\left(\int \frac{d^3p}{(2\pi)^3} F(t,t,p)\right)^2 \nonumber \\ 
&+\frac{\lambda}{12 N} \int_{t^0}^t \! dt' \! \int \! d^3 z
\Big[
- 2 I_F(t,t',|\vec{x}-\vec{z}|) F\rho(t,t',|\vec{x}-\vec{z}|) \notag \\
&\hspace{4cm} - I_\rho(t,t',|\vec{x}-\vec{z}|) \lp F^2 - \frac{\rho^2}{4} \rp (t,t',|\vec{x}-\vec{z}|)
\Big]. \label{energy}
\end{align}
In contrast to $n(t)$, $\varepsilon(t)$ should be strictly conserved, and this serves as a check of our numerical simulations. 

\section{Numerical results}\label{Sec:Numerical results}

In this section, we present the numerical solutions of Eqs.~\eqref{evolF} and~\eqref{evolR}.
We fix $N=4$ and $\lambda = 10$ throughout. We choose a rather large value of the coupling constant, $\lambda=10$, in order to achieve  thermalization on a relatively short (computing) time \cite{Hatta:2012gq}.  The initial occupancy is chosen uncorrelated to the value of the coupling constant. We  consider  three different values of $A$:  $A=0.1$, $A=1$ and $A=5$.  The results depend on two dimensionful quantities: the bare mass $m$ and the parameter $Q$ in the initial distribution function. As we shall explain shortly, we use two different values of the ratio of these parameters,  $m/Q=1/5$, or $m/Q=1/2$, depending on the value of $A$. 


  The calculations are done on a regular lattice in momentum space. We call $a$ the lattice spacing and $\Lambda$ the maximum value of the momentum on the grid. We choose the lattice spacing as $a_p=0.2\,m$, and the time step as $a_t = 0.01/m$. In order to ensure energy conservation, the momentum cutoff $\Lambda$  must be at least several times larger than $Q$, and this requirement becomes more and more severe as $A$ is increased (the final, equilibrium, distribution extends to larger momenta as $A$ grows). We use $\Lambda=2\,Q=10\,m$ (50 grid points) for (a) $A=0.1$ and (b) $A=1$, and $\Lambda=7.5\,Q=15\,m$ (75 grid points) for (c) $A=5$.  Note that, according to Fig.~\ref{ac},  the case $A=0.1$ is `underpopulated' and the cases $A=1$ and $A=5$ are `overpopulated'.  The $A=5$ case is numerically more demanding; in order to make the ratio $\Lambda/Q$ large, we increased the number of grid points and chose a larger ratio $m/Q=1/2$. 

For the calculation of the memory integrals  in Eqs.~\eqref{evolF} and~\eqref{evolR}, we exploit the fact that the self energies $\Sigma_F(t,t'')$ and $\Sigma_\rho(t,t'')$ decay rapidly as $t-t''$ increases, and we introduce a cutoff $t_c$ on the $t''$ integration:
 $\int^t_{t_0} dt^{\prime\prime} \dots\longrightarrow \int^{t}_{t-t_c} dt^{\prime\prime} \dots$ with $t_c = 50/Q$.  We have checked that the results are unchanged even if we use the smaller value $t_c=25/Q$.   
The convolution integrals in the collision term is somewhat simplified by using the formula
 \begin{align}
\int \frac{d^3k}{(2\pi)^3} A(|\vec{p}-\vec{k}|)B(k)
=\frac{1}{32\pi^2 p}\int_p^\infty du \int_{-p}^p dv (u^2-v^2)A\left(\frac{u-v}{2}\right)B\left(\frac{u+v}{2}\right).
\end{align}
%
%
Finally, we perform a mass renormalization in order to eliminate the most severe ultraviolet divergences. These appear in the  $p$-integral in Eq.~(\ref{p}), which is quadratically divergent due to the vacuum contribution to $F$. At each time step, we perform a simple subtraction and use the renormalized mass 
\begin{align}
M^2_\text{ren}(t) = m^2 + \frac{\lambda (N+2)}{6N} \int \frac{d^3p}{(2\pi)^3} (F(t,t,p) - F(0,0,p))\,.
\end{align}
The same mass renormalization is applied to the total energy (\ref{energy}). Thanks to this and our choice of the cutoff $\Lambda$, the total energy is conserved to better than 1\% accuracy in all the results to be presented below. There exist also logarithmic divergences that could be eliminated by a coupling constant renormalization. However these are milder and in practice do not affect the results \cite{Arrizabalaga:2004iw}.

Finally, we observe that large numbers are generated by a combination of the coupling constant and various factors of $\pi$ emerging from angular integrations. As a very rough estimate, after rescaling all dimensionful quantities by appropriate factors of $Q$, one  extracts, in the right hand side of Eqs.~(\ref{evolF}) and (\ref{evolR}) an overall factor $\lambda^2 Q^2/(72 N\pi^4)$ or $\lambda^2 Q^2/(1152N\pi^6)$ depending on how one estimates the angular integrals\footnote{In Eqs.~(\ref{evolF}) and (\ref{evolR}), we estimate the magnitude of $\Sigma$ as follow. In Eq.~(\ref{Sigma}) we isolate  the factor $\lambda/3N$, and in Eq.~(\ref{if}) the factor $\lambda/6$, while $F(t,t',p)$ is taken to be of order $Q^{-1}$. Collecting these factors together with factors $1/(2\pi^2)$, or $1/(8\pi^3)$, coming from the angular integrations, one obtains the estimates mentioned in the text.}. We could absorb these factors in a ``natural'' time scale $\tau$, with $t\approx 16 \,\tau$ or $t\approx 210\, \tau$ (in units $Q^{-1}$). These numbers are in fact underestimates, as we shall see, but they indicate that we should expect the dynamics to develop over time units that 
are typically in the range $100 \,Q^{-1}$ or even larger. 
\iffigure
\begin{figure}[bth]
 \includegraphics[width=8cm,height=7cm]{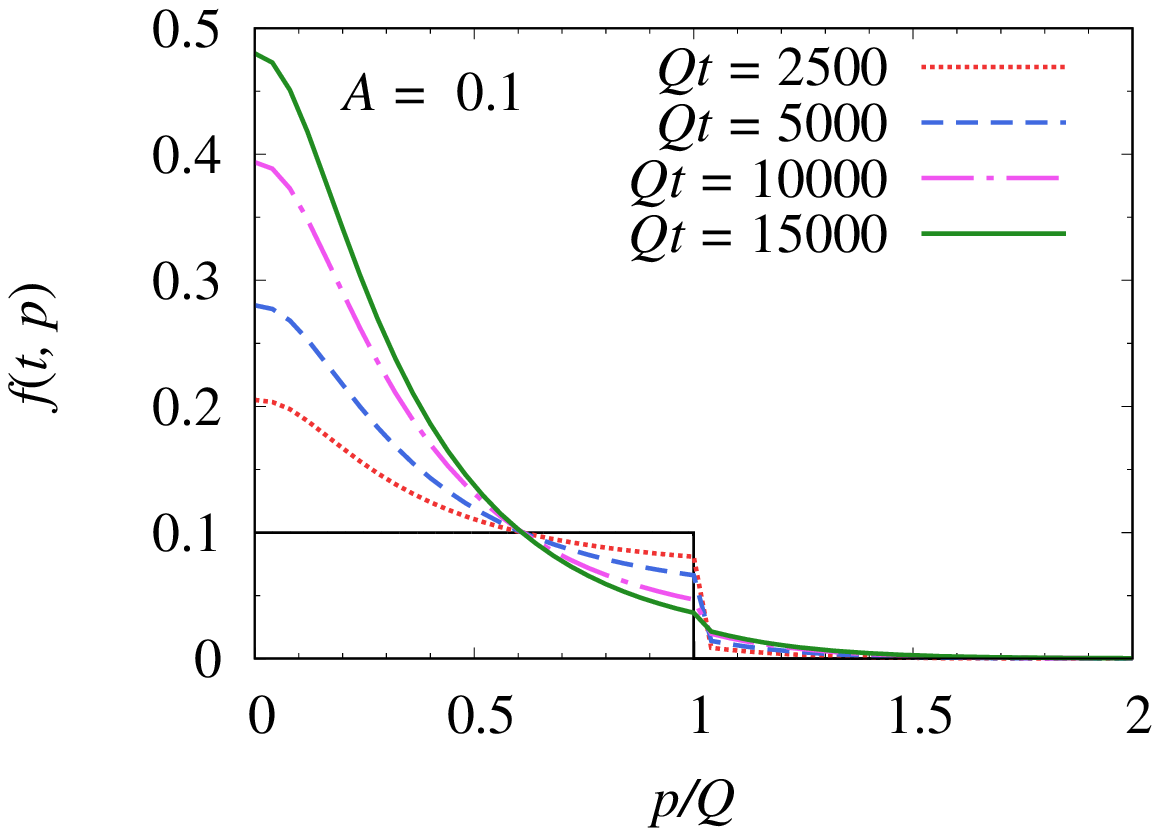}
 \includegraphics[width=8cm,height=7cm]{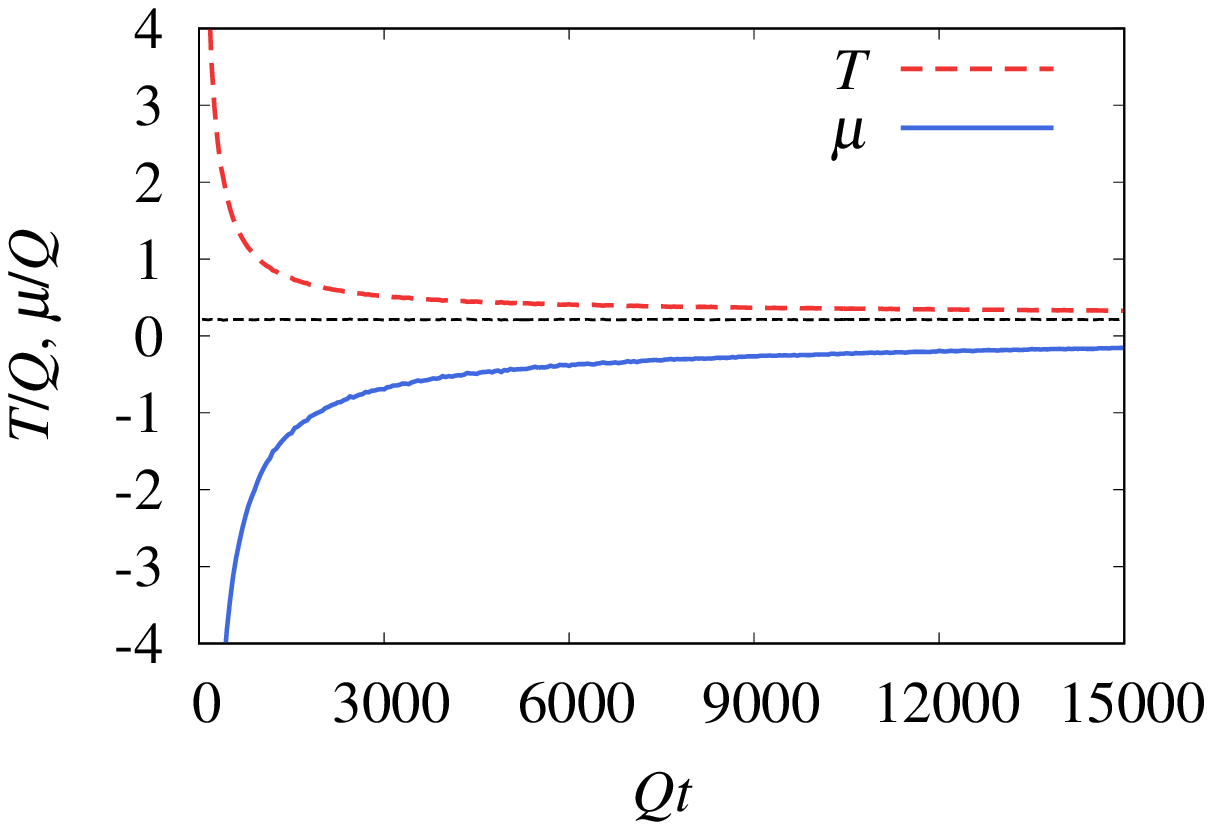}
 \includegraphics[width=8cm,height=7cm]{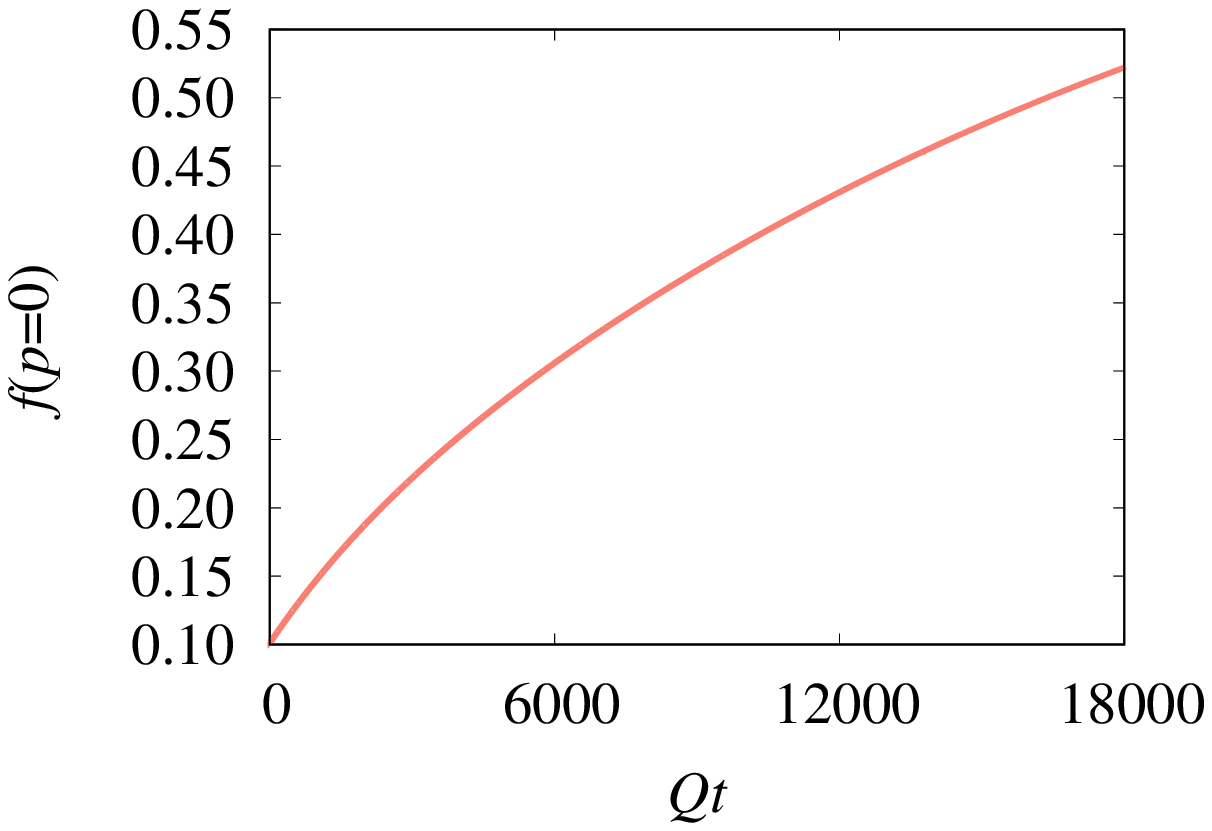}
\vspace{10mm}
\caption{
 Results for $A=0.1$. Upper left: time evolution of the particle distribution. The initial step function distribution is also shown. Upper right: fit to the Bose-Einstein distribution with finite $T$ (upper, red line) and $\mu$ (lower, blue line). The black dashed line in the middle denotes the renormalized quasiparticle mass Eq.~(\ref{mass}), which is nearly constant, $m_\text{qp}(t)\simeq Q/5$. Bottom: the occupancy of the zero mode $p=0$ as a function of time. 
	} \label{0.1} 
\end{figure}
\fi

\subsection{$A=0.1$:  underpopulated case }

We start our discussion with the underpopulated case, i.e., with $A=0.1$. The results are summarized in Fig.~\ref{0.1}. The upper left figure shows the time evolution of the distribution $f(t,p)$, and the right figure shows the temperature $T(t)$ and the chemical potential $\mu(t)$ in units of $Q$ extracted by fitting $f(t,p)$ to the Bose-Einstein distribution in the soft, low-$p$ ($p\lesssim 0.5 Q$), region. We find that such a fit is possible already at early times.  The smooth evolution of the distribution function, as well as the  behaviors of the temperature and chemical potential as a function of time, clearly show the trend towards thermalization. However, it takes a very long time for the system to fully thermalize. This can be seen by noticing that the effective chemical potential $\mu$ is always negative and eventually approaches, very slowly, the equilibrium value $\mu=0$ from below. Also, as shown in the lower figure, the zero mode $p=0$ occupancy grows monotonously  and has not reached it equilibrium value at the end of the simulation. The growth of soft modes is mostly due to elastic scattering. In fact the number density (\ref{number}) is approximately conserved: it only changes by less than 2\% during the time span of the simulation. 


\iffigure
\begin{figure}[h]
 \includegraphics[width=8cm,height=7cm]{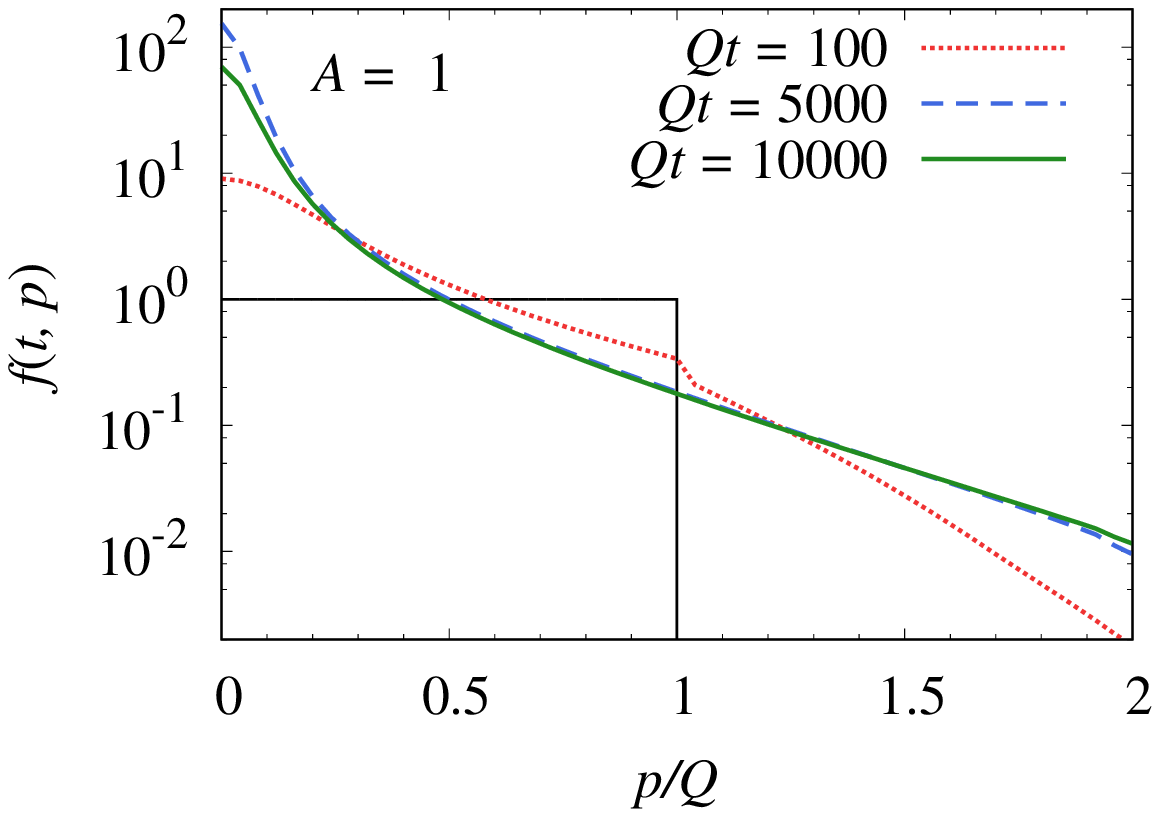}
\includegraphics[width=8cm,height=7cm]{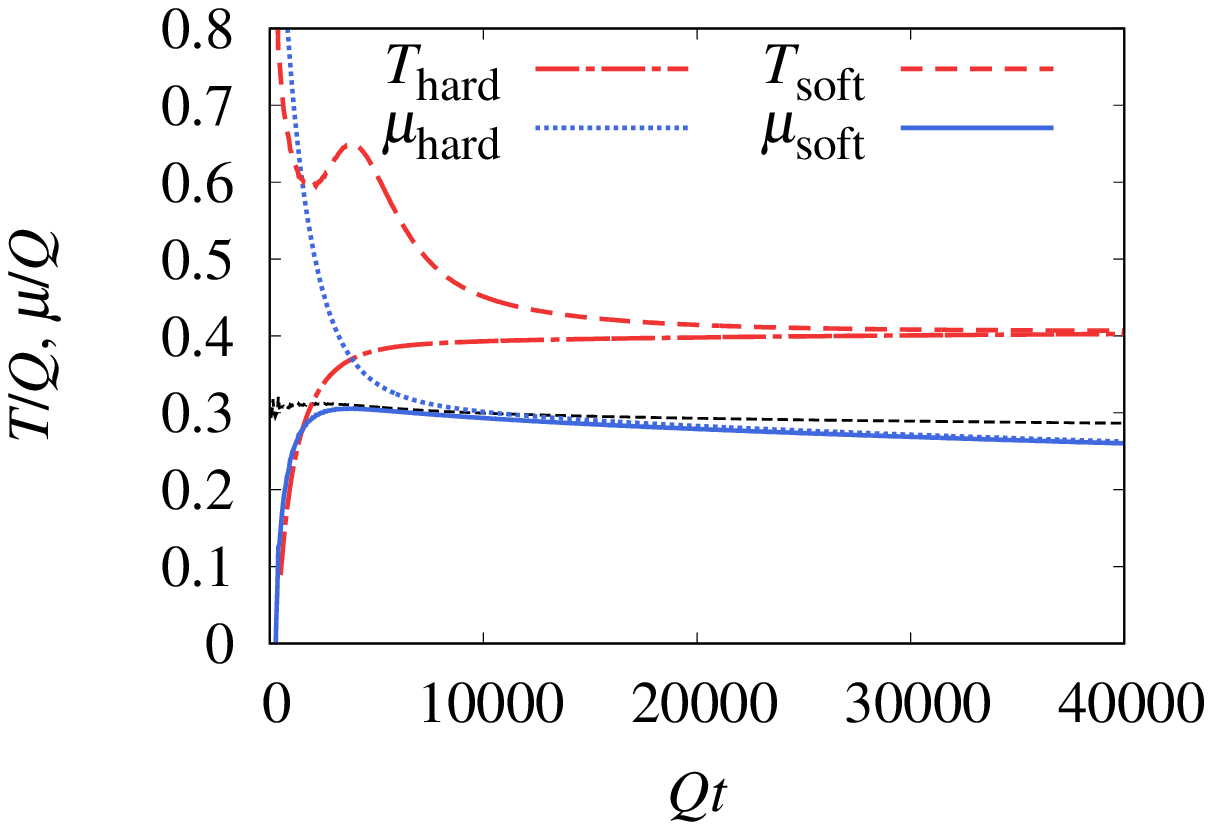}
 \includegraphics[width=8cm,height=7cm]{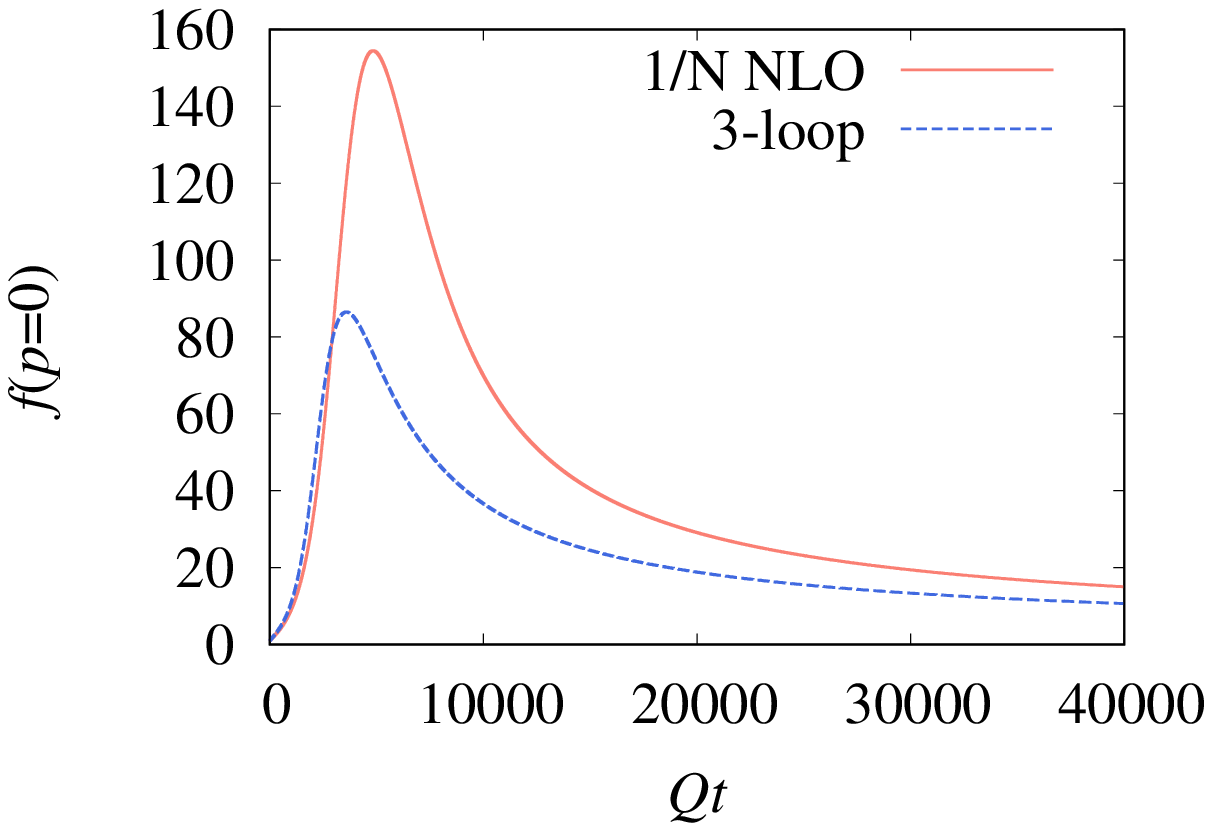}
\includegraphics[width=8cm,height=7cm]{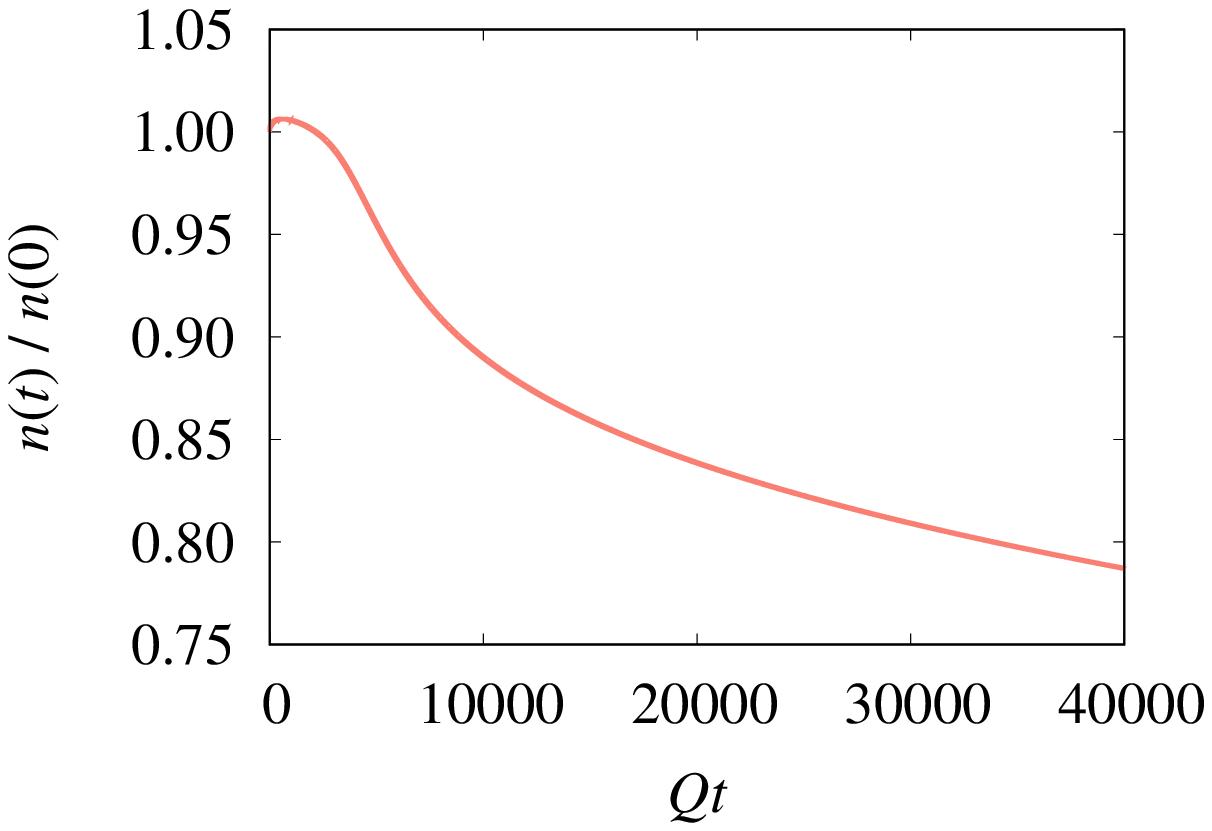}
\vspace{10mm}
\caption{
 Results for $A=1$. Upper left: the time evolution of the particle momentum distribution.  Upper right: the parameters $\mu$ (lower blue lines) and $T$ (upper red lines) obtained from a fit to the Bose-Einstein distribution in the hard  (solid lines) and soft (dotted lines) momentum regions, respectively. The black, nearly horizontal line, represents the effective mass. Lower left: time evolution of the zero momentum mode occupancy in the NLO $1/N$ (red, solid) and 3-loop (blue, dashed) cases. Lower right: total particle number, Eq.~(\ref{number}), as a function of time.
	} \label{1} 
\end{figure}
\fi

\subsection{$A=1$: overpopulated case }

Next we turn to the `overpopulated' case $A=1$, with the main results displayed in  Fig.~\ref{1}. One sees both quantitative and qualitative differences as compared to the $A=0.1$ case. The lower left figure shows that the zero mode occupancy $f(p=0)$ rises sharply at early times, reaches a maximum and then decreases slowly. The particle number is approximately conserved during the growth of the soft modes, and it starts to decrease appreciably at a later time as shown in the lower right figure. By the end of the simulation, the system has lost about 20\% of particles. Another visualization of the growth of soft modes is given  in the right figure of Fig.~\ref{qp1} where we plot the particle flux. Before $f(0)$ reaches a maximum, the flux in the soft region is positive, meaning that particles are flowing into the soft momentum region. After the peak, the flux turns negative everywhere. 

As already pointed out, the growth of soft modes is primarily due to elastic processes, which conserve particle number. Eventually inelastic processes, presumably $2$ to $4$ processes\footnote{We have no way at this point to isolate precisely which inelastic processes dominate. In leading order both $2\to4$ and $1\to 3$ (and their reverse) can contribute. However the $1\to 3$ process can only occur off shell, which is made possible in a heat bath by the thermal width acquired by the quasiparticles. In contrast $2\to 4$ processes can occur on shell, and for this reason we believe these to be the dominant ones.},  take over and start to eliminate the particles in excess in order to reach the equilibrium distribution. The existence of this ``delay'' in the action of the inelastic processes is also observed in calculations based on kinetic equations \cite{Blaizot:2016iir}. A further perspective on this feature is provided by the comparison of the full $1/N$ NLO  calculation with a 3-loop 2PI calculation where we keep only the first line of (\ref{if}) and (\ref{irho}). The corresponding occupancy of the zero momentum mode is shown in Fig.~\ref{1}, lower left panel. One sees that the inelastic collisions suppress the growth of the low momentum modes sooner than in the full NLO calculation. This we interpret as a result of the screening of the interaction in the full calculation \cite{Berges:2008wm}, making inelastic processes less effective.

In the upper right figure, we determined the set $(T,\mu)$ from the soft ($p\lesssim 0.5 \,Q$) and hard ($p\gtrsim 1.0\, Q$) momentum regions separately. 
The two determinations converge at late times, indicating that the soft and hard modes  approach thermalization. However, it takes an extremely long time (much longer than the simulation time) for the system to reach the true equilibrium state where $\mu_\text{soft}=\mu_\text{hard}=0$.  What is interesting is that, unlike what happens in  the $A=0.1$ case, where the chemical potential remains negative, here  $\mu_\text{soft}$  quickly turns positive and approaches very close to (but never exceeds) the quasiparticle mass $m_\text{qp}$ defined by Eq.~(\ref{mass}). 
 This is more clearly shown in the left plot of Fig.~\ref{qp1} (magnified version of the upper-right figure in Fig.~\ref{1}) where $\mu_\text{soft}$ and $m_\text{qp}$ differ only by less than 2\% when they are the closest, which roughly occurs at the time when $f(p=0)$ peaks out. 
This is reminiscent of the dynamical BEC formation in number-conserving theories \cite{Semikoz:1994zp,Blaizot:2013lga,Epelbaum:2014mfa,Meistrenko:2015mda}, and in fact, the particle number is approximately conserved up to this point.  The spectrum in the infrared is well approximated by the Bose-Einstein, or Rayleigh-Jeans distribution
\beq
f(t,p) \approx \frac{T(t)}{\omega_p - \mu(t)}\,,
\eeq
but does not exhibit any obvious scaling behavior \cite{Semikoz:1994zp,Berges:2015ixa}.  
Moreover, the subsequent behavior is very different from what it would be if the threshold for BEC was crossed. Since $\mu$ never becomes equal to the quasiparticle mass,  $f(p=0)$ does not diverge. Instead, it starts to decrease, and at around the same time, the total particle number $n(t)$  also starts to decrease as a result of inelastic processes, as already discussed. Such a `decay of a condensate' has also been observed in  \cite{Berges:2012us} where $f(p=0)$ exhibits a plateau behavior before starting to decrease. The reason we do not see such a plateau here is presumably due to  the strength of the coupling which makes  the recombination process more efficient.

\iffigure
\begin{figure}[bth]
 \includegraphics[width=8cm,height=7cm]{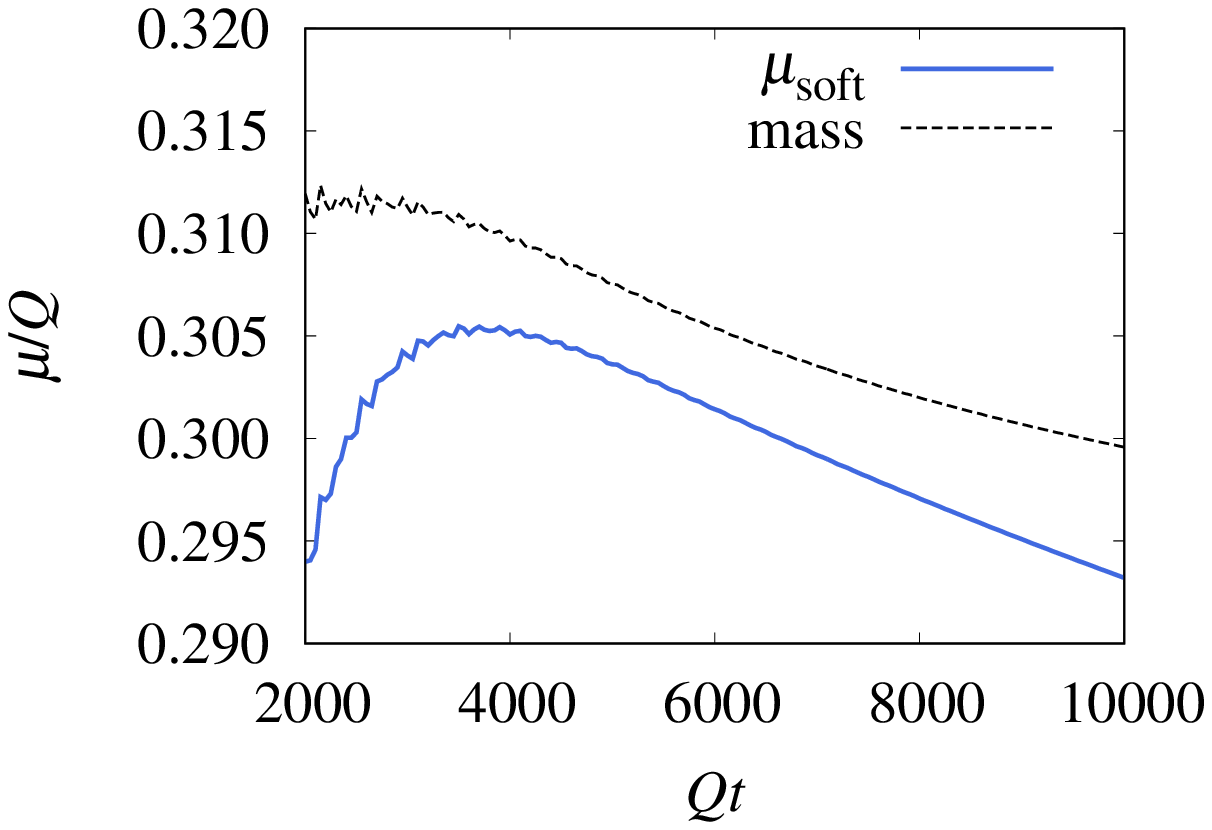}
 \includegraphics[width=8cm,height=7cm]{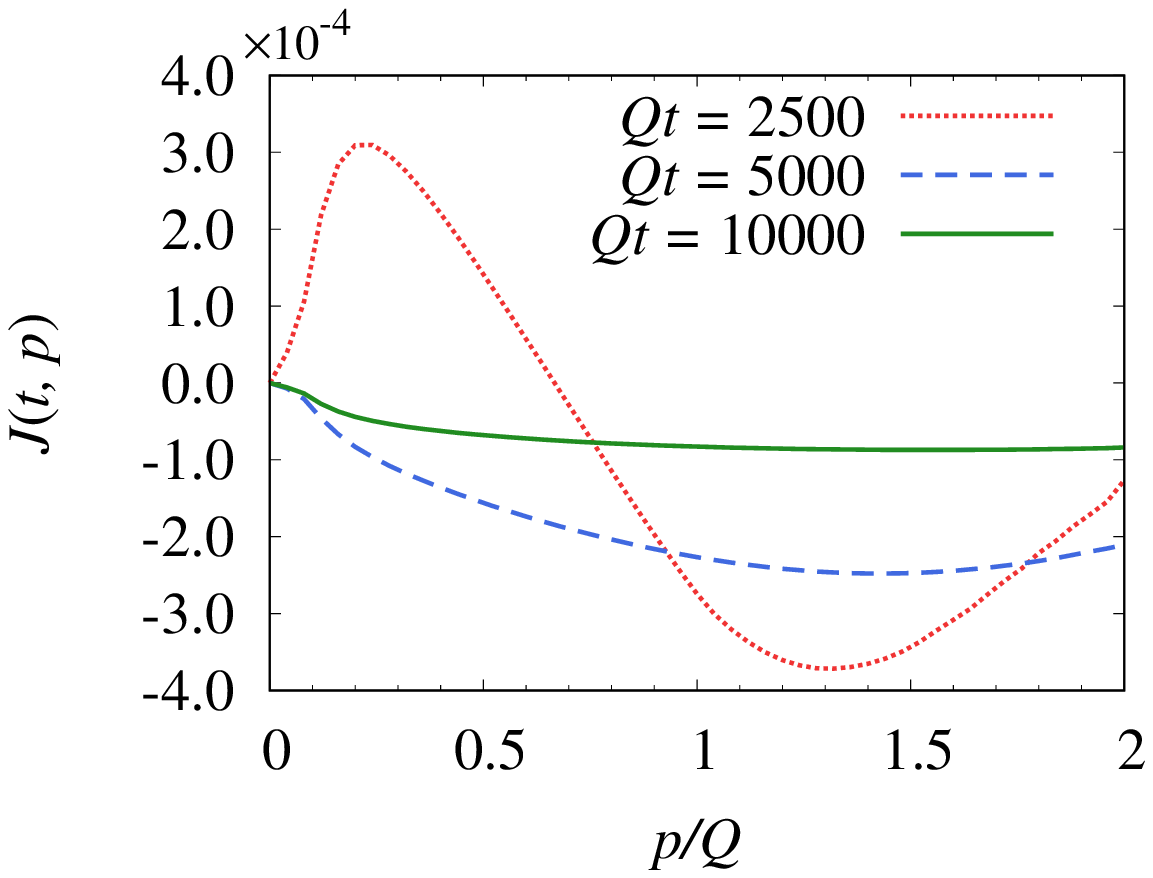}
\vspace{10mm}
\caption{
 Results for $A=1$. Left: the effective chemical potential in the soft region,  and the quasiparticle mass $m_\text{qp}$, as a function of time. Right: particle flux $J(t,p)$, Eq.~(\ref{flux}), before ($Qt=2500)$ and after ($Qt=5000, 10000$) the peak in $f(p=0)$ (see Fig.~\ref{1}). 
	} \label{qp1} 
\end{figure}
\fi

\subsection{$A=5$: strongly overpopulated case}

The results for  $A=5$, shown in Fig.~\ref{5}, do not differ much, qualitatively, from those of the previous case  $A=1$. Most comments made for this case could be repeated here. 
 The zero mode occupancy $f(p=0)$ rises more sharply, and faster,  and  it peaks out earlier than in the $A=1$ case. 
 At around the peak, $\mu_\text{soft}$ makes its closest  approach to  $m_\text{qp}$.   After that, both $f(p=0)$ and the total particle number start to decrease, the latter dropping by about 40\% by the end of the simulation. Given that the value $A=5$ is safely labeled `overpopulated', these results corroborate our discussion above, that the formation of a  BEC is hindered by inelastic processes. It is interesting to note that the competition between elastic and inelastic processes forces the system to spend a fair amount of time in the vicinity of the onset  of condensation, a situation somewhat reminiscent of that of non thermal fixed points. 

\iffigure
\begin{figure}[bth]
 \includegraphics[width=8cm,height=7cm]{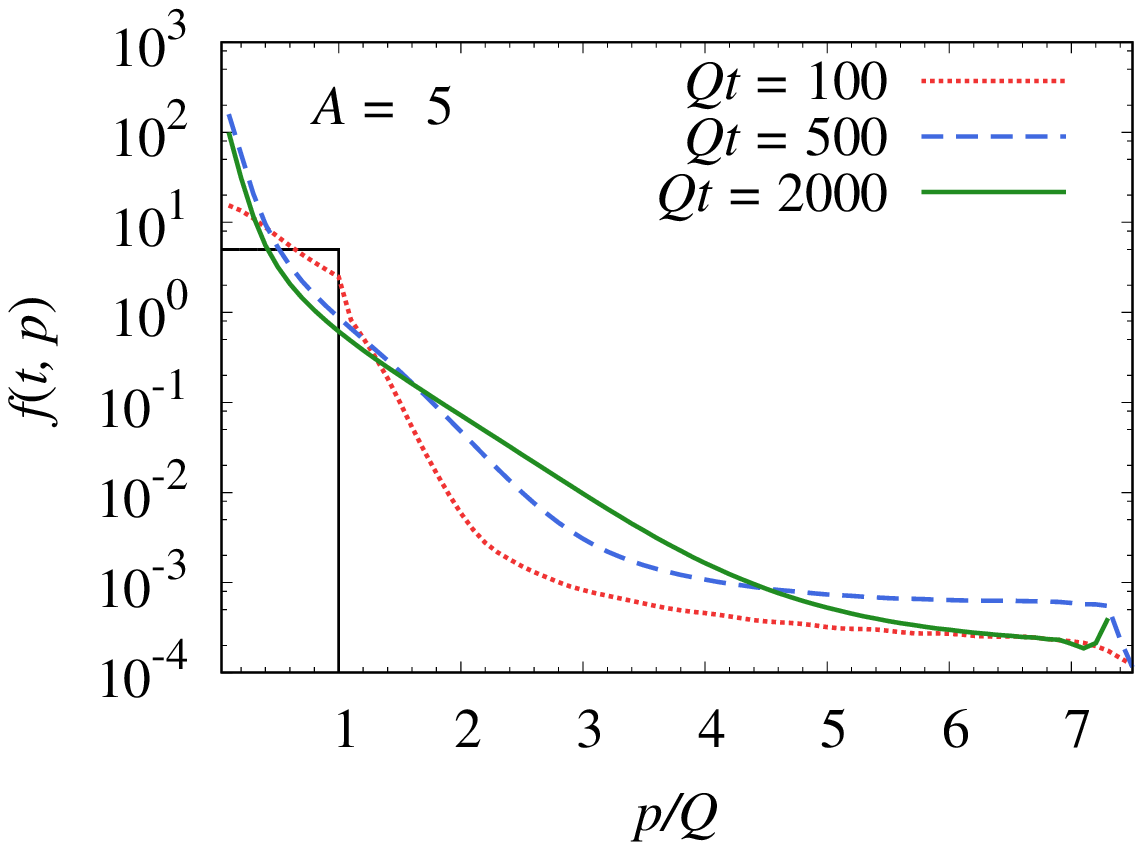}
\includegraphics[width=8cm,height=7cm]{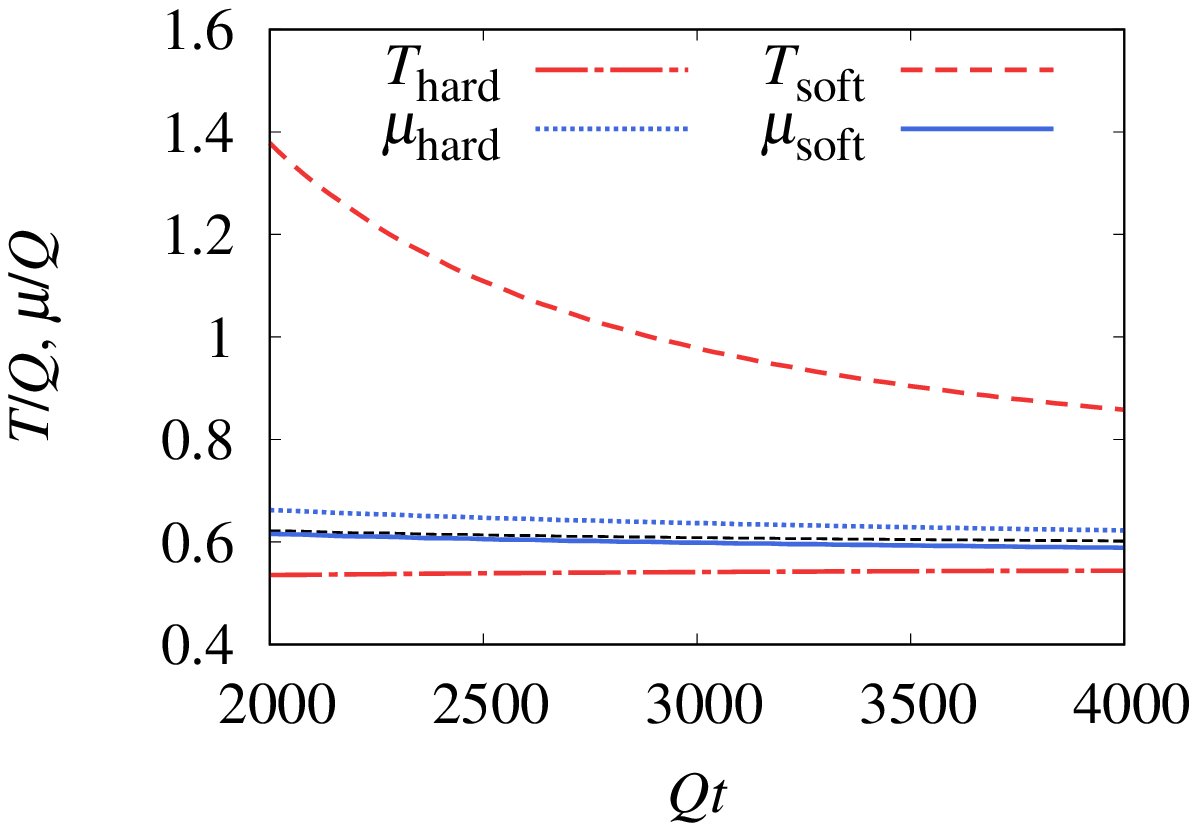}
 \includegraphics[width=8cm,height=7cm]{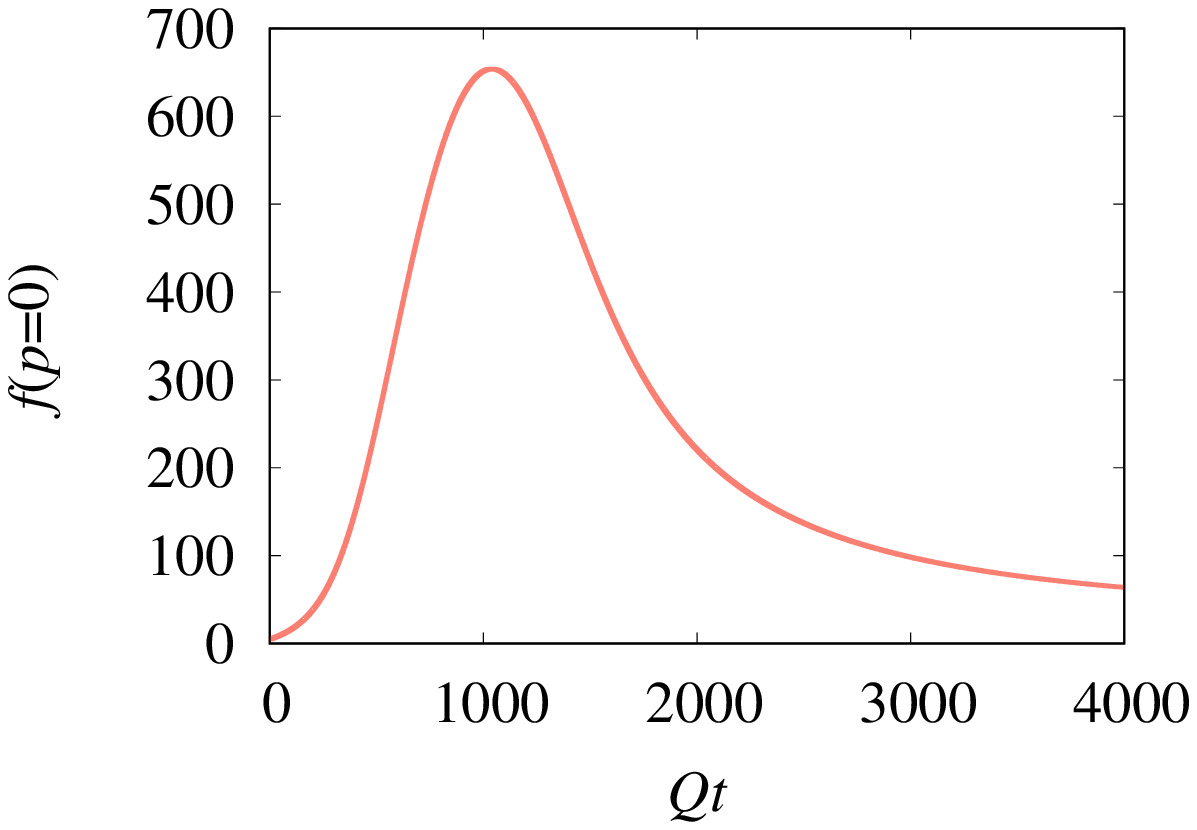}
\includegraphics[width=8cm,height=7cm]{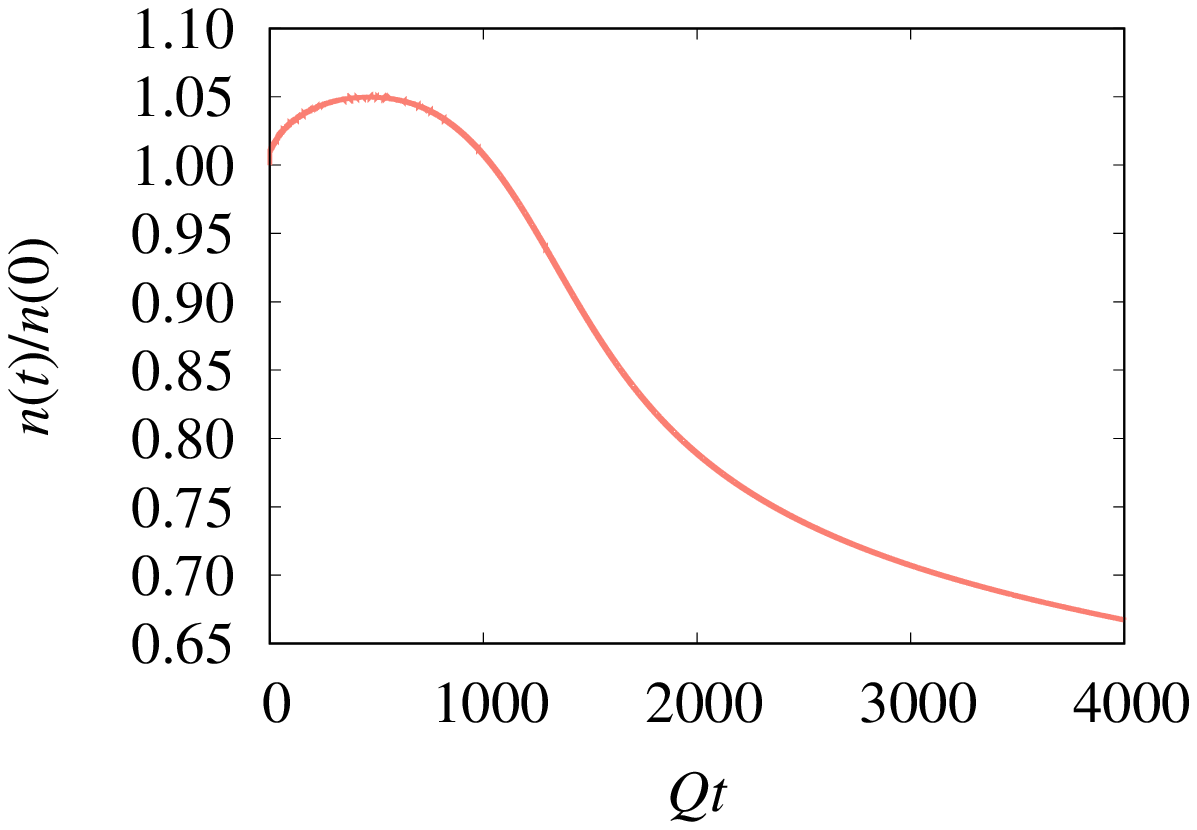}
\vspace{10mm}
\caption{
 Results for $A=5$. The description of the figures are the same as in Fig.~\ref{1}. 
	} \label{5} 
\end{figure}
\fi

\iffigure
\begin{figure}[bth]
 \includegraphics[width=8cm,height=7cm]{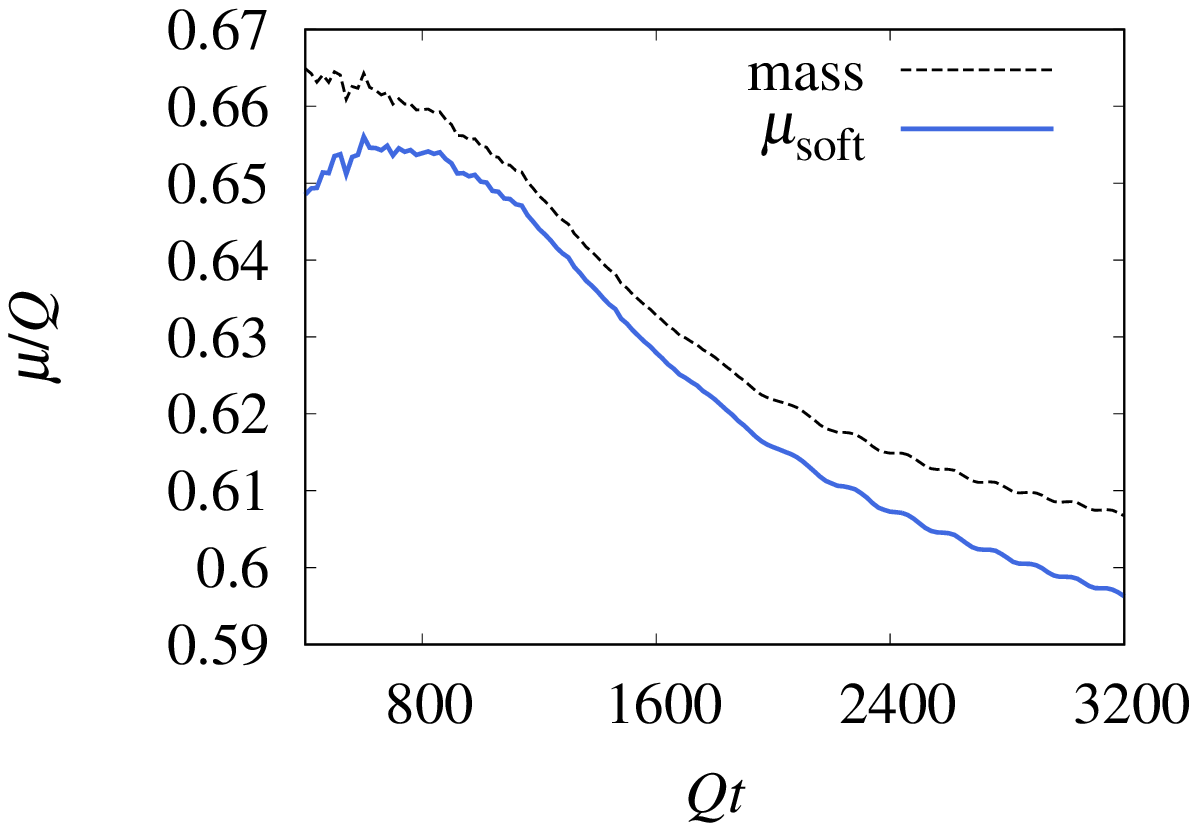}
 \includegraphics[width=8cm,height=7cm]{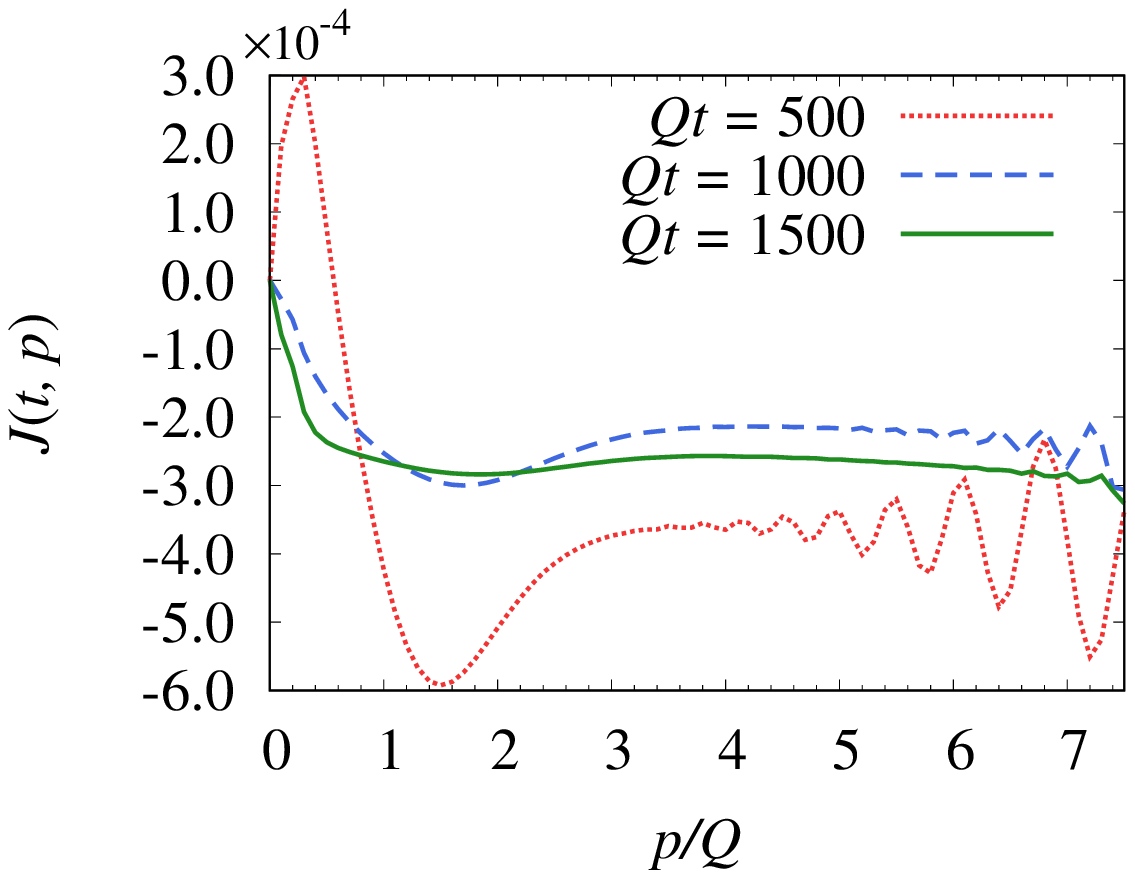}
\vspace{10mm}
\caption{
	Results for $A=5$. Left: the effective chemical potential in the soft region,  and the quasiparticle mass $m_\text{qp}$, as a function of time. Right: particle flux $J(t,p)$, Eq.~(\ref{flux}), before ($Qt=500)$ and after ($Qt=1000, 1500$) the peak in $f(p=0)$ (see Fig.~\ref{5}). The oscillations at large $p$ at early times is because $J\sim \partial_t f(t,p) \sim \partial_t (F \omega_p(t))$, and the dispersion relation $\omega_p(t)$ oscillates at large $p$ at such early times.  } \label{qp5} 
\end{figure}
\fi

\section{Discussions and conclusions}\label{Sec:conclusion}

Due to technical reasons, it is difficult to go beyond $A=5$ in our present numerical setup. Since our lattice size is limited, already when $A=5$ we had to employ a  $Q$ value smaller than in the $A=1$ case in order to make the ratio $\Lambda/Q$ large and ensure  cutoff-independence. The case $A=10$, $\lambda=10$ has been studied in the same formalism in \cite{Berges:2016nru}, and the evolution of $f(t,p)$ reported there is qualitatively similar to the $A=1$ and $A=5$ cases above.\footnote{Ref.~\cite{Berges:2016nru} used massless fields $m=0$ and observed that $f(t,p)$ approaches the Bose-Einstein distribution with $\mu= 0$. In our simulations, $m$ is kept finite, and this allows us to study in detail the transient regime where $\mu$  approaches $m$ before relaxing to the true equilibrium value $\mu=0$. } 
 Actually, in Ref.~\cite{Berges:2016nru} the authors considered the region $1000\ge A\ge 10$ with fixed $A\lambda=100$. When $A\gg 10$, which corresponds to the weak coupling regime, the spectrum can no longer be fitted by the Bose-Einstein or Rayleigh-Jeans distribution. This is a far-from-equilibrium regime where nonthermal fixed points appear and $f(t,p)$ exhibits self-similar scaling behaviors. Our study is complementary to this, being more focused on the regime where one can discuss the formation of a BEC on top of  the Bose-Einstein distribution.


In the previous studies of the BEC formation based on the Boltzmann equation, special care has been taken in order to treat the singular behavior near the condensation onset. For instance, Ref.~\cite{Semikoz:1994zp} used a much larger lattice with logarithmic spacings in $p$ to allow soft particles to decay into even softer ones, instead of accumulating them in a single $p=0$ mode as in our simulation. Besides, usually one splits the Boltzmann equation into two equations, one for the zero mode and the other for nonzero modes. It is not clear to us whether such more elaborate treatments can qualitatively change the overall picture that emerges from our calculations.  Our result suggests, at least for the range of parameters that we have considered, that the formation of a BEC is hindered by  the particle number changing processes. Amusingly though, condensation is approached, and the competition between elastic and inelastic processes forces the system to spend a relatively long time in the vicinity of the condensation onset, an effect somewhat reminiscent of that of non-thermal fixed points observed for other ranges of parameters.       

It would be interesting to extend the present analysis to longitudinally expanding systems as in heavy-ion collisions.
The BEC formation in the presence of longitudinal expansion has been studied in \cite{Epelbaum:2015vxa,Berges:2015ixa} in different approaches. The NLO-2PI simulation in the longitudinally expanding geometry was performed in 
 \cite{Hatta:2012gq} but in 1+2 dimensions (see also \cite{Tranberg:2008ae}). It should be straightforward to generalize the work of  \cite{Hatta:2012gq} to 1+3 dimensions.  

Finally,  applications of the 2PI formalism to nonequilibrium phenomena in QCD have been so far quite limited, and there are only a few attempts in the literature which  however take into account only low-order 2PI diagrams \cite{Nishiyama:2010mn,Hatta:2011ky}. Unfortunately, a systematic all-order resummation scheme of 2PI diagrams (like the $1/N$ expansion in the scalar theory) is lacking in gauge theory.

\section*{Acknowledgments}
We thank Akira Ohnishi for many discussions. This work was initiated when J.-P.~B. was a visiting professor at Yukawa Institute for Theoretical Physics, Kyoto University.
S.T. is supported by the Grant-in-Aid for JSPS fellows (No.26-3462). J.-P.~B thanks Fran\c{c}ois Gelis and J$\ddot{\rm u}$rgen Berges for useful discussions.




\end{document}